\documentclass[a4paper,reqno]{article}
\usepackage{amsmath,amssymb,amstext}
\usepackage[body={15cm,20.5cm}]{geometry}
\usepackage{url}
\usepackage{graphicx}
\usepackage{subfigure}
\graphicspath{{growth-figs/}}

\begin{document}

\begin{center}

{\Large \bf On Harmonic Measure of the Whole-Plane Levy-Loewner Evolution}

\vspace{3mm}

Igor \textsc{Loutsenko}

\vspace{3mm}

Centre de Recherches Math\'ematiques, Universit\'e de Montr\'eal\\
e-mail: loutseni@crm.umontreal.ca\\[1mm]

\vspace{3mm}

Oksana \textsc{Yermolayeva}

\vspace{3mm}

Institut Henri Poincar\'e, UPMC Paris VI\\
e-mail: yermolay@ihp.jussieu.fr\\[1mm]


\vspace{5mm}

Abstract

\begin{quote}
Generalizing our results in exact description of multifractal spectrum of the whole-plane SLE,
we consider a class of radial Levy-Loewner evolutions and compute exactly sets of points in their average means beta-spectrum.

\end{quote}

\end{center}

\section{Introduction}

We start with a simple introduction to radial Levy-Loewner evolution
(a good introduction to the chordal LLE can be found in
\cite{ORGK}, \cite{ROKG}, for a quick introduction to Levy processes see e.g. \cite{A} and references therein).

Let us consider iterative conformal mappings $z=F_n(w)$ from the
exterior of the unit disc in the $w$-plane to the exterior of a
bounded, simply connected domain in the $z$-plane: The $n$th mapping
is a composition of $n$ elementary ``spike" mappings $z=f_i(w,
\delta t_i)$, $i=1..n$
\begin{equation}
F_{n}(w)=F_{n-1}(f_n(w, \delta t_n)), \quad F_0(w)=w,
\label{Fn}
\end{equation}
where
\begin{equation}
f_n(w, t)=e^{\i\varphi_n}h\left(e^{-\i\varphi_n} w, t\right), \quad
h(w, t)=e^{ t}(w+1)\frac{w+1+\sqrt{(w+1)^2-4e^{- t}w}}{2w}-1 \label{atom}
\end{equation}
The elementary mapping $z=f_n(w, \delta t_n)$ attaches a radial
``spike" of the length $\sqrt{\delta t_n}\left(1+ O(\delta
t_n)\right)$  located at the angle $\varphi_n$ to unit disc:
Here, the point $w=e^{\i\varphi_n}$ on the unit circle in the
$w$-plane is mapped to the tip of the spike in the $z$-plane (see
Figure 1).

The mapping $z=h(w, t)$ that attaches a spike to the disc at
$\varphi=0$ satisfies the simplest Loewner equation $ \frac{\partial
h(w, t)}{\partial t}=w\frac{\partial h(w, t)}{\partial
w}\frac{w+1}{w-1} $ and as a consequence
$$
\frac{\partial f_n(w, t)}{\partial t}=w\frac{\partial f_n(w,
t)}{\partial w}\frac{w+e^{\i\varphi_n}}{w-e^{\i \varphi_n}}, \quad f_n(w,0)=w.
$$
This equation is invariant wrt any conformal transformation $f_n(w,
t)\to F\left(f_n(w, t)\right)$ and therefore an iterative compound
mapping $F_n$ in (\ref{Fn}) can be represented as a solution of the
Loewner equation at time $t=\sum_{i=1}^n \delta t_i$:
$$
F_n(w)=F\left(w, \sum_{i=1}^n \delta t_i\right), \quad F(w,0)=w,
$$
\begin{equation}
\frac{\partial F(w, t)}{\partial t}=w\frac{\partial F(w,
t)}{\partial w}\frac{w+e^{\i L(t)}}{w-e^{\i L(t)}} \label{LE}
\end{equation}
where $L(t)$ is the piecewise constant function
$$
L(t)=\varphi_i, \quad \sum_{j=1}^{i-1} \delta t_j < t < \sum_{j=1}^i
\delta t_j
$$
One can consider continuous-time Loewner evolutions as limits of
the above iterative processes when $n\to\infty$ and $\delta t_n\to
0$. The iterative picture will be useful for the simplest
derivation of linear integro-differential equation for moments of
derivatives of conformal mappings for the Levy-Loewner evolution by
the method proposed by M. Hastings \cite{H}.

\begin{figure}
 \hspace{6.5cm}\includegraphics[width=2.5cm]{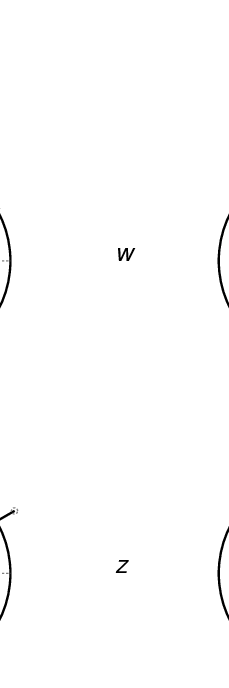}
 \caption{\small The elementary mapping $z=f_n(w, \delta t_n)$ (left) and the compound iterative mapping $z=F_n(w)$ (right) from exterior of unit circle in $w$-plane to exterior of a simply connected domain in $z$-plane.}
 \end{figure}
We are interested in the case when $L(t)$ is a stochastic process
without drift. Without loss of generality we set
\begin{equation}
L(0)=0, \quad \langle L(t) \rangle = 0,
\label{Drift=0}
\end{equation}
where $\langle \rangle$ denote expectation (ensemble average).

When a stochastic process $L(t)$ is continuous in time, the conformal
mapping $z=F(w,t)$ describes growth of a random continuous curve
$\Gamma=\Gamma(t)$ starting from a point on a unit circle $|z|=1$ at
$t=0$. On the other hand, when $L(t)$ is discontinuous in time, the
growth branches in the $z$-plane.

If one requires that the Loewner evolution (\ref{LE}) is a
conformally invariant Markovian process, in the sense that the time
evolution is consistent with composition of conformal maps \footnote{which implies that the probability distribution of $z=F^{-1}\left(F(w,t+\tau),t\right)$ coincides with that of $z=F(w,\tau)$, where $w=F^{-1}(z,t)$ is an inverse of $z=F(w,t)$},
then the necessary condition for such an evolution is that $L(t)$ must have independent stationary
increments, i.e. $L(t)$ is a Markovian process with the probability distribution of
$L(t+\tau)-L(t)$ depending only on $\tau$.
Among such processes there is a class of Levy processes considered in the present paper.

The only continuous (modulo uniform drift) process of Levy type is
the Brownian motion
\begin{equation}
L(t)=B(t)
\label{B}
\end{equation}
The Brownian motion is characterized by a single parameter -
``temperature" $\kappa$ :
\begin{equation}
\langle (B(t+\tau)-B(t))^2 \rangle = \kappa |\tau| . \label{Bt}
\end{equation}

Stochastic Loewner evolution driven by Brownian motion is called
Schramm-Loewner Evolution (SLE, or SLE$_\kappa$). Since it describes
non-branching planar stochastic curves with conformally-invariant
probability distribution, SLE is a useful tool for description of
boundaries of critical clusters in two-dimensional equilibrium
statistical mechanics. In this picture, different $\kappa$
correspond to different classes of models of statistical mechanics
(a good introduction to SLE for physicists can be found e.g. in \cite{C}, \cite{G} as
well as mathematical reviews can be found e.g. in \cite{GL},
\cite{GL2}).

By now, SLE is well studied with many exact results obtained. In particular, SLE is the only non-trivial example
where the multi-fractal spectrum as well as related $\beta$-spectrum of
the curve is described exactly.

In this article we would like to present new results in study of multi-fractal spectrum of SLE/LLE. Mainly, these results concern the ``unbounded" version of the whole-plane LLE/SLE, where we exactly determine points of integral means $\beta$-spectra for certain classes of Levy processes. The studies of such processes were initiated by B.Duplantier et al in \cite{DNNZ}. In our previous works \cite{L}, \cite{LY} we proposed an effective approach to the problem based on derivation of exact solutions of equations for moments of derivatives of conformal mappings.

The $\beta$-spectrum is a Legendre transform of the dimension spectrum that quantitatively describes subsets of the domain boundary where certain scaling laws apply (see Appendix 1)
\footnote{For further introduction to multi-fractal analysis see e.g.
\cite{BS}, \cite{HJKPS}, \cite{H}, \cite{Wiki} and references therein} to derivatives
of conformal mappings $F'(w)$ along the boundary.
The integral means $\beta(q)$-spectrum of the
domain is defined through the $q$th moment of derivative at the unit
circle (i.e. at $|w|\to 1$) as follows
\begin{equation}
\beta(q)=\overline{\lim}_{\epsilon\to0+}\frac{\log
\int_0^{2\pi}\langle\left|F'\left(e^{\epsilon+\i\varphi}\right)\right|^q\rangle
d\varphi}{-\log \epsilon} \label{beta}
\end{equation}
This spectrum can be written down explicitly for the whole-plane SLE and several points of the spectrum can be found for certain types of the whole-plane LLE.

Studies of exact multi-fractal properties for the "bounded" (see below) version of the whole-plane SLE has been performed in \cite{BS}. The Loewner evolution driven by different Levy processes and analysis of their spectra were performed by several authors. For example, they appear in \cite{JS} for the case of "bounded" radial LE driven by compound Poisson process and in \cite{CR} the chordal LE is studied for symmetric $\alpha$-stable processes. In both cases spectra are trivial.

The ``bounded" whole-plane LLE is a properly scaled infinite-time limit of the radial LLE
\begin{equation}
\mathcal{F}(w,t)=\lim_{T\to\infty}e^{-T}F(w,T+t),
\label{wp+ext}
\end{equation}
which describes growth of stochastic curve out of the point of
origin in the plane. Obviously, $\mathcal{F}(w,t)$ also satisfies the
radial Loewner equation(\ref{LE}) with time running from $t=-\infty$ to $t=\infty$.

The ``unbounded" version of the whole-plane LLE is an
inversion $\mathcal{F}(w,t)\to 1/\mathcal{F}(1/w,t)$ of the bounded version. Obviously, this mapping from the unit disc in the $w$-plane to the complement of stochastic curve which grows from infinity towards the origin in the $z$-plane satisfies the Loewner equation (\ref{LE}). It has been
studied mainly due to its relationship with the problem of Bieberbach
coefficients of conformal mappings for such processes \cite{DNNZ}, \cite{L}:
Expectations of the squares of the
Bieberbach coefficients as well as two non-trivial points of the
$\beta$-spectrum have been conjectured for two special examples of
unbounded whole-plane LLEs in \cite{DNNZ}.
The proof of this conjecture has first been proposed by one of the authors in \cite{L}
for the first
example and partially for the second one. In this article we complete the proof of this conjecture.

Note that the partial (and unsuccessful) proof of the second case has been attempted in \cite{DNNZ1}. This proof is written for Levy processes whose first $N$ characteristic coefficients coincide with those of Brownian motion. Proof of non-trivial cases involves $N>1$. Unfortunately, on the circle \footnote{This is different from the case on line, where e.g. addition of special $2\pi \mathbb{Z}$ jumps \cite{DNNZ1} could extend processes with the required properties to non-Brownian motion. Since we are considering the radial evolution, a combination of Brownian motion with the $2\pi \mathbb{Z}$ "jumps" is still a circular Brownian motion (see also section 4)}, the Levy processes possessing such a property for $N>1$ and that are different from the Brownian motion do not exist.

In the sequel we also sketch a derivation of the full $\beta(q)$ spectrum
\begin{equation}
\beta(q)=\overline{\lim}_{\epsilon\to0+}\frac{\log
\int_0^{2\pi}\langle\left|\mathcal{F}'\left(e^{-\epsilon+\i\varphi}\right)\right|^q\rangle
d\varphi}{-\log \epsilon}
\label{unbounded_beta}
\end{equation}
for the ``unbounded" whole plane SLE.

It is important to note that study of the unbounded version of the whole-plane SLE turns out to be useful not only in relation with the coefficient problem, but also for complete derivation of the version-independent multi-fractal spectrum of the bulk of SLE: The attempt of rigorous derivation the bulk spectrum (by D. Belyaev and S. Smirnov \cite{BS}) based on the bounded version only happens to be incomplete (see e.g. Appendix of \cite{LY1}).

\section{Equation for Moments of Derivative}

 Let us consider ``exterior" radial LLE, i.e. mapping from exterior of the unit disc in the $w$-plane to the exterior of the simply-connected domain in the $z$-plane (as shown in Figure 1). As has been mentioned above the bounded version of the whole-plane LLE is the properly scaled infinite-time limit of this process. So, let us call the bounded version as the ``exterior problem".\\

The derivation for the unbounded version, which is referred as the ``interior problem" is similar.\\

To find $\beta(q)$-spectrum (\ref{beta}) one needs to estimate
moments of derivative
$\langle\left|F'\left(w,t\right)\right|^q\rangle$.

It is now convenient to change the variable $w$ to
$e^{\i L(t)}w$ and consider the mapping
$$
\tilde F(w,t)=F\left(e^{\i L(t)}w, t\right)
$$
for which the point $w=1$ is the pre-image of the growing tip of the
curve. Obviously,
$$
\beta(q)=\lim_{\epsilon\to0}\frac{\log \int_0^{2\pi} \tilde
\rho\left(e^{\epsilon+\i\varphi}, e^{\epsilon-\i\varphi};
t|q\right), d\varphi}{-\log \epsilon},
$$
where
\begin{equation}
\tilde \rho(w,\bar w;
t|q)=\langle\left|\tilde F'\left(w,t\right)\right|^q\rangle .
\label{rho}
\end{equation}
It turns out that the moments $\tilde \rho(w,\bar w; t|q)$ satisfy
an integral linear equation. The simplest derivation of such
equation uses M. Hasting's iterative approach \cite{H}, \cite{L}: In
iterative picture (\ref{Fn}) we consider the mappings
$$
\tilde F_{n}(w)=\tilde F_{n-1}\left(e^{\i(\varphi_n-\varphi_{n-1})}h(w,\delta
t_n)\right) .
$$
By the chain rule
$$
\left|\tilde F'_{n}(w)\right|^q=\left|\tilde F'_{n-1}\left(e^{\i(\varphi_n-\varphi_{n-1})}h(w,
\delta t_n)\right)\right|^q \left| h'(w,\delta t_n) \right|^q .
$$
Taking expectations of the both sides of the above equation, with
the account of (\ref{rho}), we get
$$
\tilde \rho(w, \bar w, t_{n-1}+\delta t_n)=\left| h'(w,\delta t_n)
\right|^q\int_0^{2\pi}d\varphi P(\varphi, \delta t_n)\tilde
\rho\left(e^{\i \varphi}h(w,\delta t_n), e^{-\i \varphi}\bar h(\bar
w,\delta t_n) , t_{n-1}\right) ,
$$
where $P(\varphi,t)$ is the probability density that $L(t)=\varphi$ under
condition that $L(0)=0$.

Taking into account that for small $\delta t$
$$
h(w, \delta t)=w+w\frac{w+1}{w-1}\delta t ,
$$
in the first order in $\delta t$ we obtain
$$
\tilde \rho(w, \bar w, t+\delta t)=\left(1+\delta t
\left(w\frac{w+1}{w-1}\partial_w+\bar w\frac{\bar w+1}{\bar
w-1}\partial_{\bar w}- \right.\right.\qquad\qquad\qquad\qquad\qquad
$$
$$
\qquad\qquad\qquad\qquad\qquad\left.\left.-\frac{q}{(w-1)^2}-\frac{q}{(\bar
w-1)^2}+q\right)\right) \int_0^{2\pi} P(\varphi, \delta
t)\tilde\rho\left(e^{\i\varphi}w, e^{-\i\varphi}\bar w, t\right)
d\varphi ,
$$
so, in the $\delta t\to 0$ limit we arrive at the equation for
$\tilde\rho$
\begin{equation}
\partial_t \tilde\rho=\left(-\hat\eta+w\frac{w+1}{w-1}\partial_w+\bar w\frac{\bar w+1}{\bar w-1}\partial_{\bar w}-\frac{q}{(w-1)^2}-\frac{q}{(\bar w-1)^2}+q\right)\tilde\rho .
\label{Ltrho}
\end{equation}
Here the operator $\hat \eta$ acts on functions of $w, \bar w$ as follows
$$
\hat\eta[\rho](w, \bar w)=\lim_{t\to
0}\frac{1}{t}\int_0^{2\pi}\left(\rho(w, \bar
w)-\rho\left(e^{\i\varphi}w, e^{-\i\varphi}\bar
w\right)\right) P(\varphi, t) d\varphi . 
$$
We consider only the processes for which the above limit exists. On the unit circle, these Levy processes are defined by
their characteristic exponents $\eta_m$
\footnote{Since $L(t)$ is Markovian and $L(t+\tau)-L(t)$ depends only on $\tau$, $\left<e^{iuL(t+\tau)}\right>=\left<e^{iuL(t)}\right>\left<e^{iu(L(t+\tau)-L(t))}\right>=\left<e^{iuL(t)}\right>\left<e^{iuL(\tau)}\right>$, and therefore $\left<e^{iuL(t)}\right>$ must have an exponential form $\left<e^{iuL(t)}\right>=e^{-t\eta(u)}$ }
$$
e^{-t\eta_m}=\langle e^{\mathrm{i} m L(t)}\rangle=\int_0^{2\pi}e^{\mathrm{i} m \varphi} P(\varphi, t) d\varphi 
$$
(i.e. $e^{-t\eta_m}$ is the Fourier transform of the probability distribution of $L(t)$).

According to the above definitions, the operator $\hat\eta$ acts diagonally on the basis of
two-dimensional Taylor/Laurent expansions $w^n\bar w^m$, $n,m\in
\mathbb{Z}$
\begin{equation}
\hat\eta[w^n\bar w^m]=\eta_{n-m}w^n\bar w^m .
\label{hat_eta_w_w}
\end{equation}
Since we consider processes without drift,
the characteristic exponents are real and symmetric
$$
\eta_m=\bar \eta_m, \quad \eta_m=\eta_{-m}, \quad \eta_0=0 .
$$
In the case of  Brownian motion (\ref{Bt}), $P(\varphi,t)$ is a
fundamental solution of the heat equation on the circle and
\begin{equation}
\eta_m=\frac{\kappa m^2}{2}, \quad
\hat\eta=\frac{\kappa}{2}\left(w\partial_w-\bar w\partial_{\bar
w}\right)^2,
\label{eta_kappa}
\end{equation}
i.e. equation (\ref{Ltrho}) becomes the second order linear PDE.

\section{The Whole Plane LLE}

Equation (\ref{Ltrho}) simplifies in the case of the whole plane LLE
which is a properly scaled infinite-time limit of the radial LLE (\ref{wp+ext}).

Consider now the Laurent expansion of
$\tilde{\mathcal{F}}(w,t)=\mathcal{F}\left(e^{\i L(t)}w,t\right)$ at
$w=\infty$
$$
\tilde{\mathcal{F}}(w,t)=e^{t+\i L(t)}\left(w+\sum_{i=0}^\infty
\Phi_i(t)w^{-i}\right)
$$
According to the above equation and (\ref{rho})
\begin{equation}
\tilde\rho=e^{qt}\langle \left| 1-\sum_{i=1}^\infty i\Phi_iw^{-i-1} \right|^q  \rangle =e^{qt}\rho, \quad \rho=\sum_{i=-1, j=-1}^\infty
\frac{\rho_{i,j}}{w^{i+1}\bar w^{j+1}}, \quad \rho_{-1,-1}=1
\label{rho_Laurent}
\end{equation}
By existence of limit (\ref{wp+ext}),
the probability distribution of $e^{-t-\i
L(t)}\tilde{\mathcal{F}}(w,t)=w+\sum_{i=0}^\infty
\Phi_i(t)w^{-i}$ is time-independent
\footnote{For details on existence of this infinite-time limit see Section 4.3 of \cite{L} or Section 1.2 of \cite{BS}}.
So, in the case of
the whole plane LLE, the function $\rho=e^{-qt}\tilde\rho=\langle \left| 1-\sum_{i=1}^\infty i\Phi_iw^{-i-1} \right|^q  \rangle$ does not
depend on time and
$$
\frac{\partial \tilde\rho}{\partial t}=q\tilde\rho.
$$
Therefore, for the whole-plane LLE, equation (\ref{Ltrho}) becomes
\begin{equation}
L\rho=q\rho \label{Lrho}
\end{equation}
\begin{equation}
L=-\hat\eta+w\frac{w+1}{w-1}\partial_w+\bar w\frac{\bar w+1}{\bar
w-1}\partial_{\bar w}-\frac{q}{(w-1)^2}-\frac{q}{(\bar w-1)^2}+q
\label{Leta}
\end{equation}
To find moments of derivatives we have to look for a non-vanishing
and analytic at $w=\infty$ solution of the above equation. Note that such a solution is (up to a constant factor)
unique.

Indeed, by analyticity of conformal mapping at $w=\infty$, the
function $\rho(w, \bar w)$ is analytic at infinity and
$\rho(\infty)=1$. Then substituting the expansion
(\ref{rho_Laurent}) into (\ref{Lrho}) we get the nine-term
``two-dimensional" recurrence relation for the expansion
coefficients $\rho_{ij}$
\begin{equation}
\sum_{l=0}^{2}\sum_{m=0}^2C^{lm}_{ij}\rho_{i-l,j-m}=0 \label{rec9}
\end{equation}
with the following boundary conditions (see (\ref{rho_Laurent}))
$$
\rho_{-1,-1}=1, \quad \rho_{i,j<-1}=\rho_{j<-1,i}=0
$$
and the recurrent coefficients (note the symmetry $C_{ij}^{lm}=C_{ji}^{ml}$)
$$
C_{i,j}^{0,0}=-\eta_{i-j}-i-j-2 , \quad C_{i,j}^{1,1}=-4\eta_{i-j} ,
\quad C_{i,j}^{2,2}=-\eta_{i-j}+i+j-2-2q ,
$$
$$
C_{i,j}^{0,1}=2\left(\eta_{i-j+1}+i+1\right), \quad
C_{i,j}^{0,2}=-\eta_{i-j+2}+j-i-2-q, \quad
C_{i,j}^{1,2}=2\left(\eta_{i-j+1}-j+1+q\right) .
$$
Any element $\rho_{kn}$ can be found in a consecutive manner
starting from $(i,j)=(-1,-1)$ and going along the ``row" up to $j=n$
and repeating this procedure for consecutive rows up to $i=k$ by
expressing $\rho_{ij}$ as a linear combination of 8 elements:
$\rho_{i-l,j-m}, l\in \{0,1,2\}, m\in \{0,1,2\}, (l,m)\not=(0,0)$.
This fixes $\rho(w,\bar w)$ uniquely. Therefore, once an analytic
and non-vanishing at $w=\infty$ solution of (\ref{Lrho}) is found,
it will correspond to the moments of derivatives.

Recall that in the exterior problem (i.e. bounded whole-plane LLE) considered above, the curve grows
from the origin towards infinity. In the interior problem (i.e. unbounded
whole-plane LLE), curve grows from the infinity
towards the origin. Both versions are related by the inversion $
\mathcal{F}(w,t)\to 1/\mathcal{F}(1/w,t)$. Obviously, the unbounded
mapping satisfies the same Loewner equation (\ref{LE}), but now
$\mathcal{F}$ maps the interior of the unit disc to the complement of the
unbounded curve in the plane
$$
\mathcal{F}(w,t)=e^{-t}\left(w+\sum_{i=2}^\infty \mathcal{F}_i(t)
w^i\right), \quad |w|<1
$$
By analogy with the exterior problem, the moments of derivatives
$$
\tilde\rho=\langle |\mathcal{F}'(e^{iL(t)}w,t)|^q\rangle
$$
depend on time only through the exponential scaling, i.e. (note the
difference in the exponent sign in comparison with
(\ref{rho_Laurent}))
$$
\tilde \rho=e^{-qt}\rho ,
$$
where
\begin{equation}
\rho(w,\bar w; q)=\sum_{i=1,j=1}^\infty \rho_{ij}w^{i-1}\bar
w^{j-1}, \quad \rho_{1,1}=1 \label{rho_ij_int}
\end{equation}
is time-independent.

Similarly to the exterior problem, $\rho(w,\bar w;q)$ satisfies an
integral equation which now writes as follows
\begin{equation}
L\rho=-q\rho,
\label{Lrho_int}
\end{equation}
where $L$ is given by (\ref{Leta}). An analytic and
non-vanishing at $w=0$ solution of (\ref{Lrho_int}) corresponds now to
the moments of derivatives of the interior problem.

The expansion coefficients of the interior problem $\rho_{ij}$ (see Eq.(\ref{rho_ij_int})) satisfy the recurrence relation of the type (\ref{rec9}) with the following boundary conditions
$$
\rho_{11}=1, \quad \rho_{i<0, j}=\rho_{i, j<0}=0
$$
and recurrence coefficients ($C_{ij}^{lm}=C_{ji}^{ml}$)
$$
C_{i,j}^{0,0}=-\eta_{i-j}-i-j+2 , \quad
C_{i,j}^{1,1}=-4(\eta_{i-j}-2q) , \quad
C_{i,j}^{2,2}=-\eta_{i-j}+i+j-6+2q ,
$$
$$
C_{i,j}^{0,1}=2\left(\eta_{i-j+1}+i-1-q\right), \quad
C_{i,j}^{0,2}=-\eta_{i-j+2}+j-i-2+q, \quad
C_{i,j}^{1,2}=2\left(\eta_{i-j+1}+3-j-2q\right)
$$

\section{Universal Points, $q=2$}

It has been observed by B. Duplantier et al in \cite{DNNZ} that in the interior problem $\rho_{ii}=i^2$, $ \beta=3$, when $q=2$ and the evolution is driven by any symmetric Levy process with $\eta_1=3$. The second observation consists in that for $q=2$ and evolution driven by any symmetric process with $\eta_1=1$ one has $\rho_{ii}=i^3$, $\beta=4$.

The first observation has been proved in \cite{L} using the coincidence of SLE and LLE solutions of Eq. (\ref{Lrho_int}) when $q=2$ and $\eta_1=3$. The SLE sub-case of the second observation has been also proved in \cite{L}. Below we present the complete proof of conjectures by B. Duplantier et al containing both cases
\footnote{ The proof of sub-case of the second observation with $\eta_1=1, \eta_2=4$ has been also presented in \cite{DNNZ1}. Unfortunately, it has been unnoticed by authors of \cite{DNNZ1} that, in contrast with the Levy processes on line, such symmetric Levy processes on circle (i.e. $L(t)$ modulo $2\pi\mathbb{Z}$) different from the Brownian motion do not exist. This fact follows from elementary analysis of Levy-Khinchin formula for characteristic exponents on circle}:\\

{\it\bf Theorem}:  Let us take the interior whole plane Loewner evolution and fix $q=2$, then:

\begin{enumerate}

\item For evolution driven by a symmetric Levy
processes with  $\eta_1=3$
$$
\rho_{ii}=i^2, \quad \beta(2)=3
$$
\item For processes with $\eta_1=1$
$$
\rho_{ii}=i^3, \quad \beta(2)=4
$$

\end{enumerate}

Proof:\\

Representing
$\rho(w, \bar w)$ in the form
\begin{equation}
\rho(w, \bar w)=(1-w)(1-\bar w)\Theta(w, \bar w)
\label{rho1}
\end{equation}
from (\ref{Lrho_int}) with $q=2$ we get
\begin{equation}
-\hat\eta[(1-w)(1-\bar w)\Theta]+(w+1)(\bar w
-1)w\frac{\partial\Theta}{\partial w}+(\bar w+1)(w -1)\bar
w\frac{\partial\Theta}{\partial \bar w}+3(2w \bar w-w-\bar
w)\Theta=0 \label{Ltheta1}
\end{equation}
where $\Theta$ is the series
\begin{equation}
\Theta=\theta_0(\xi)+\sum_{i=1}^{\infty}\left(w^i+\xi^i w^{-i}
\right) \theta_i(\xi), \quad \xi=w\bar w \label{theta_1}
\end{equation}
Substituting it into (\ref{Ltheta1}) we will get a three-term
differential recurrence relation for $\theta_i(\xi)$.

Indeed, making the change of variable $\bar w=\xi/w$, then taking
into account the facts that, according to (\ref{hat_eta_w_w}),
operator $\hat \eta$ commutes with $\xi=w\bar w$ and that $\hat \eta
[w^i]=\eta_i w^i$, after substitution of series (\ref{theta_1}) into
(\ref{Ltheta1}) we will get a series in $w$ with coefficients
depending on $\xi$. Coefficients of $w^i$ define the
following three-term differential recurrence relations
\begin{equation}
2\xi(\xi-1)\theta'_i(\xi)-\left(\eta_i+i+(\eta_i-i-6)\xi\right)\theta_i(\xi)+\xi(\eta_i+i-2)\theta_{i+1}(\xi)+(\eta_i-i-2)\theta_{i-1}(\xi)=0, \quad \theta_{-i}(\xi)=\xi^i\theta_i(\xi)
\label{rec_diff}
\end{equation}
It is easy to see that this recurrence relation truncates at $i=N$, i.e. $\theta_i=0$ for $i \ge N$, when $\eta_N=N+2$ for some $N$.

The simplest truncation happens when $N=1$, i.e. when $\eta_1=3$, which corresponds to the case 1) of the Theorem. In this case only $\theta_0$ does not vanish identically and due to (\ref{rec_diff})
$$
(\xi-1)\theta'_0(\xi)+3\theta_0(\xi)=0
$$
Therefore $\theta_0(\xi)=1/(1-\xi^3)$, and $\rho(w, \bar w)=(1-w)(1-\bar w)/(1-w\bar w)^3$. Then using definitions of $\rho_{i,j}$ and $\beta(q)$ (see Eqs. (\ref{rho_ij_int}) and (\ref{unbounded_beta}) respectively) we prove the case 1) of the Theorem.\\

Let us now consider the case 2) of the Theorem, i.e. $q=2$, $\eta_1=1$. Now the truncation of the recurrence relation does not take place. However, it turns out that in this case, two equations defining functions $\theta_0(\xi)$ and
$\theta_1(\xi)$ do not involve others $\theta_i(x), i\not=0, 1$.

In the case 2) of the Theorem equations for $i=0$ and $i=1$ take the form
$$
(\xi-1)\theta_0'(\xi)+3\theta_0(\xi)-2\theta_1(\xi)=0, \quad
\xi(\xi-1)\theta_1'(\xi)+(3\xi-1)\theta_1(\xi)-\theta_0(\xi)=0
$$
Since function $\rho$ is analytic at $w=0$, as well as
$\rho(w=0)=1$, we have to look for a solution of the above system of two ODEs,
such that $\theta_0(0)=1$ and $\theta_1(\xi)$ is finite at $\xi=0$.
With these conditions we get
$$
\theta_0=\frac{1+\xi}{(1-\xi)^4}, \quad
\theta_1=-\frac{1}{(1-\xi)^4}
$$
Again, taking definitions of $\rho_{i,j}$ and $\beta(q)$ into account we prove the case 2) of the Theorem.

\section{$\beta(q)$-Spectrum of the Interior Whole-Plane SLE}

We recall that in the case (\ref{B}), (\ref{Bt}), when the driving Levy process is the Brownian motion, the Loewner evolution represents the growth of a continuous curve (SLE-curve).  In the interior whole-plane SLE the curve has two exceptional singular points: One of them is the moving tip ot the curve, and the second is an immovable "tip" at infinity.
Equations for moments of derivatives (\ref{Lrho_int}), or (\ref{Lrho}) takes form of the parabolic PDE of the second order with
\begin{equation}
L=-\frac{\kappa}{2}\left(w\partial_w-\bar w\partial_{\bar
w}\right)^2+w\frac{w+1}{w-1}\partial_w+\bar w\frac{\bar w+1}{\bar
w-1}\partial_{\bar w}-\frac{q}{(w-1)^2}-\frac{q}{(\bar w-1)^2}+q
\label{L_kappa}
\end{equation}
The $\beta$-spectrum of the curve is so far determined by combination of two methods: For $q$ which is smaller than the critical value $q<Q(\kappa)$, where
\begin{equation}
Q(\kappa)=\frac{\kappa^2+8\kappa+12-2\sqrt{2\kappa^2+16\kappa+36}}{16\kappa} ,
\label{Q}
\end{equation}
the blow-up rates of any positive solution of equation (\ref{Lrho_int}, \ref{L_kappa}) turn out to depend only on singularity of solution in vicinity of the moving tip of the curve, i.e. at $w=1$: Near the moving tip, $\epsilon\to 0$, $w=1+\i \epsilon \zeta$, $\bar w=1-\i\epsilon\bar \zeta$, the equation (\ref{Lrho_int}, \ref{L_kappa}) reduces to
$$
\left(-\frac{\kappa}{2}\left(\frac{\partial}{\partial \zeta}+\frac{\partial}{\partial \bar \zeta}\right)^2+\frac{2}{\zeta}\frac{\partial}{\partial\zeta}+\frac{2}{\bar\zeta}\frac{\partial}{\partial \bar\zeta}-q\left(\frac{1}{\zeta^2}+\frac{1}{\bar \zeta^2}\right)\right)\rho(\zeta,\bar\zeta)=0,
$$
The latter has a solution of the form
$$
\rho=(\zeta-\bar \zeta)^{-\kappa\gamma^2/2}(\zeta\bar\zeta)^\gamma
$$
where
\begin{equation}
\gamma(q,\kappa)=\frac{\kappa+4-\sqrt{(\kappa+4)^2-8q\kappa}}{2\kappa}
\label{gamma}
\end{equation}
A part of the $\beta$-spectrum is then determined by the blow-up rates of such a solution, i.e. by exponents $\gamma$ and $-\kappa\gamma^2/2$.

This property of positive solutions has been used by D.Belyaev and S.Smirnov in \cite{BS} for the rigorous calculation of the part of the $\beta$-spectrum of the exterior whole-plane SLE
\footnote{The idea of using a PDE for moments of derivatives and singularities of its solution to determine multi-fractal spectrum of SLE first appeared in work by M. Hastings \cite{H}}.
Their approach relies on maximum principle for parabolic PDE. In the case of the interior problem it works only for $q<Q(\kappa)$. Therefore, for bigger values of $q$ one has to use different methods. A method based on exact solutions of (\ref{Lrho_int}, \ref{L_kappa}) has been proposed by us in \cite{LY, LY1}.

In more details: Representing an analytic at $w=0$ solution of (\ref{Lrho_int}, \ref{L_kappa}) in the form
\begin{equation}
\rho(w, \bar w)=\left(\left(1-w\right)\left(1-\bar w\right)\right)^\gamma\Theta(w, \bar w), \quad q=2\gamma+\frac{1}{2}\kappa\gamma-\frac{1}{2}\kappa\gamma^2,
\label{rho_to_theta}
\end{equation}
where $\gamma$ is given by (\ref{gamma}) and function $\Theta$ has the following expansion around $w=0$
\begin{equation}
\Theta=\sum_{n \in \mathbb{Z}} w^n f_n(\xi), \quad \xi:=w\bar w, \quad f_{-n}(\xi)=\xi^n f_n(\xi), \quad \bar f_n=f_n
\label{whole_int}
\end{equation}
we come to the following three-term differential recurrent relation for $f_n(\xi)$:
\begin{equation}
\xi{A}_{n+1} f_{n+1} + {A}_{-n+1} f_{n-1}+\left({B}_n+(1-\xi){C}_n\right) f_{n}+2\xi(\xi-1)\frac{df_n}{d\xi}=0,
\label{rec3}
\end{equation}
with
$$
{A}_n=\frac{\kappa}{2}(n-\gamma)^2+n-3\gamma-\frac{\kappa}{2}\gamma(1-\gamma),
$$
$$
{B}_n=-\kappa (n^2+\gamma^2-\gamma)+6\gamma, \quad {C}_n=\kappa\frac{n^2-2\gamma+2\gamma^2}{2}-n-6\gamma .
$$
It follows from the recurrence relation (\ref{rec3}) that expansion (\ref{whole_int}) truncates at $|n|=M$ (i.e. $f_n=0$, for $|n|>M$) if $A_{-M}=0$. This truncation condition determines a countable number of curves in the parametric $(q, \kappa)$-plane:
\begin{equation}
\kappa=2\frac{M+3\gamma}{M^2+2M\gamma+2\gamma^2-\gamma}, \quad q=\frac{\gamma(M+\gamma)(2M+1+\gamma)}{M^2+2M\gamma+2\gamma^2-\gamma}
\label{kq_int}
\end{equation}
Along these curves, relations (\ref{rec3}) transform into a system of $M+1$ first order linear ODEs for $f_0, \dots, f_M$ (remember that $f_{-n}(\xi)=\xi^n f_n(\xi)$). The blow-up rates of solution of PDE (\ref{Lrho_int}, \ref{L_kappa}) at the unit circle $\xi=1$ are determined by asymptotic of $f_n(\xi)$ at $\xi\to 1$. In this asymptotic the analysis of system of ODEs is effectively reduced to analysis of eigenvalues of $2M+1\times 2M+1$ three-diagonal matrix. The blow-up rates then give the following $\beta(q)$ spectrum along the set of curves (\ref{kq_int}) (see \cite{LY, LY1}):
\begin{equation}
\beta=\left\{\begin{array}{ll}
\kappa\frac{\gamma^2}{2}-2\gamma-1 , &  \quad q\le -1-\frac{3\kappa}{8} \\
\kappa\frac{\gamma^2}{2}, & \quad -1-\frac{3\kappa}{8} \le q \le Q(\kappa) \\
3q-\frac{1}{2}-\frac{1}{2}\sqrt{1+2q\kappa}, & \quad  q \ge Q(\kappa)
\end{array}\right.
\label{results}
\end{equation}
where $\gamma=\gamma(q,\kappa)$ and $Q(\kappa)$ are given by (\ref{gamma}) and (\ref{Q}) respectively.

\section{Exterior Problem and Spectrum of the bulk of SLE}

The similar picture takes place in the case of the exterior problem. The approach (by D. Belyaev and S.Smirnov in \cite{BS}) based on analysis of singularities at $w=1$ and maximum principle for parabolic PDE (\ref{Lrho}, \ref{L_kappa}) works for $q>{\mathcal Q}^+(\kappa)$,
where
$$
{\mathcal Q}^+(\kappa)=-(\kappa+4)^2(\kappa+8)/128
$$
(for details see Appendix of \cite{LY1}).

Our method of derivation of the $\beta$-spectrum of the exterior problem for $q\le{\mathcal Q}^+(\kappa)$ relies on exact solutions of PDE for moments of derivative along the countable number of curves in the parametric $(q, \kappa)$-plane \cite{LY1}: By analogy with the interior problem one seeks for a  solution of PDE (\ref{Lrho}, \ref{L_kappa}) that is now analytic at $w=\infty$
$$
\rho=\left((1-w^{-1})(1-\bar w^{-1})\right)^\gamma\Theta
$$
with $\Theta$ being a Fourier polynomial of the $M$th degree
$$
\Theta(w, \bar w)=\sum_{n=-M}^M w^{- n} f_n\left(\xi\right) , \quad \xi=1/(w\bar w), \quad f_{-n}(\xi)=\xi^n f_n(\xi), \quad \bar f_n=f_n
$$
The polynomiality (i.e. truncation) condition determines countable number of curves in the parametric $(q, \kappa)$-plane:
$$
\kappa=2\frac{M-\gamma}{M^2+2M\gamma+\gamma}, \quad q=\frac{\gamma(M+\gamma)(2M+1+\gamma)}{M^2+2M\gamma+\gamma},
$$
Again, along these curves, the derivation of the $\beta$-spectrum reduces to analysis of eigenvalues of three-diagonal $2M+1\times 2M+1$ matrix \cite{LY1}. In this version of SLE the $\beta$-spectrum has the following form
$$
\beta=\left\{\begin{array}{ll}
\kappa\frac{\gamma(q,\kappa)^2}{2}-2\gamma(q,\kappa)-1 , &  \quad q\le -1-\frac{3\kappa}{8} \\
\kappa\frac{\gamma(q,\kappa)^2}{2}, & \quad -1-\frac{3\kappa}{8} \le q \le \frac{3(\kappa+4)^2}{32\kappa} \\
q-\frac{(\kappa+4)^2}{16\kappa}, & \quad  q \ge \frac{3(\kappa+4)^2}{32\kappa}
\end{array}
\right.
$$
Note, that in difference from the interior problem, in this version of SLE the transition in the spectrum does not take place at $q={\mathcal Q}^+(\kappa)$. Despite of existence of a transition in blow-up rates of solutions at the unit circle $|w|=1$ far from the moving tip $w=1$, the average $\beta$-spectrum is still determined by behavior of solution at $w\to 1$.

However, this transition in blow-up rates leads to complications in derivation of the bulk spectrum of SLE. In more details: The bulk spectrum is independent of version of SLE and defined for generic points of the curve away from the tip. For the finite-time radial evolution it is determined through the integration away from the tip $w=1$:
\begin{equation}
\beta_{\rm bulk}(q)=\overline{\lim}_{\epsilon\to0+}\frac{\log
\int_\delta^{2\pi-\delta}\langle\left|\tilde F'\left(w=e^{\epsilon+\i\varphi}\right)\right|^q\rangle
d\varphi}{-\log \epsilon}
\label{beta_bulk}
\end{equation}
On the other hand, in contrast to the finite-time evolution the exterior whole-plane SLE has two tips
: one moving tip at as well as the origin of the curve
\footnote{An immovable tip at the origin $z=0$ in the exterior problem corresponds to "tip" at $z=\infty$ of the interior problem}. The conformal mapping has additional singularity at the origin of the curve, that gives own contribution to the integral. An effect of this contribution results in the above mentioned transition in the blow-up rates away from $\varphi=0$ at $q={\mathcal Q}^+(\kappa)$ (for details see Appendix of \cite{LY1}). Therefore, definition (\ref{beta_bulk}) is not suitable for all values of $q$ in the case of the whole-plane SLE
\footnote{ The simplest example illustrating the above statement is the whole plane SLE at $\kappa$=0: The mapping in this case is $\mathcal{F}(w,t)=e^t(w+1)^2/w$. Integration out of the moving tip $w=1$ does not allow to get rid of contribution at the origin $w=-1$ and as a consequence so defined "bulk spectrum" does not differ from the tip spectrum for all values of $q$.}.

Thus, to complete the derivation of the bulk spectrum one should consider the blow-up rates away from $\varphi=0$ for both versions of the whole plane SLE: the exterior version for $q \ge 0$ and the interior version for $q \le 0$.
The Legendre transform of so derived bulk spectrum corresponds to the multi-fractal harmonic spectrum predicted from quantum gravity by B.Duplantier \cite{D}.

\section{Conclusion}

In summary: From elementary consideration it follows that any
analytic and non-vanishing at $w=\infty$/$w=0$ solution of the
equation for moments (\ref{Lrho})/(\ref{Lrho_int}) is unique (up to
a constant factor) and, therefore, provides a point in the
$\beta$-spectrum of the exterior/interior whole plane LLE,
respectively.

The analysis of exact solutions with such properties allows to compute exactly sets of points of average $\beta$-spectrum for a Loewner evolution driven by certain classes of Levy processes, as well as find complete tip and bulk $\beta$-spectra of the whole-plane SLE.




\vspace{5mm}

\noindent
{\large\textbf{Appendix 1: Harmonic measure and multifractal spectra}}\\

In this Appendix we briefly recall the connections between different spectra characterizing possibly fractal sets.

We start with the definition of $f(\alpha)$-spectrum: Suppose that some measure is defined on our curve (this can be electric flux, probability, mass, magnetization etc). Then dividing the space into boxes of size $l$, we look for the distribution of scalings of the measure wrt change of $l$ inside the boxes containing the curve.

For instance, consider the harmonic measure of a planar curve at the point $z=F(w)$, which is proportional to the probability \footnote{Which is, in turn, inversely proportional to $|F'(w)|$} of a random walker released from infinity to hit the curve at $z$. For the curve covered with $N$ boxes of size $l$, we denote the probability of hitting the curve at the $i$-th box by $p_i$.

Then as $l\to 0$
\begin{equation}
\sum_{i=1}^N p_i^q \asymp \int l^{-f(\alpha)}l^{q\alpha}d\alpha
\label{falpha}
\end{equation}
i.e. the number of boxes for which probability scales as $l^{\alpha}$ with the change of $l$ is proportional to $l^{-f(\alpha)}$.

The $f(\alpha)$-spectrum is related to observable properties of the measure through a Legendre transform. These properties are expressed through the $\tau(q)$-spectrum \footnote{It is easy to see that $-\tau(0)$ is the Hausdorff dimension of the curve. Another common definition is generalized dimension $D(q)=\tau(q)/(q-1)$ with $D(0)$, $D(1)$ and $D(2)$ being the Hausdorff, information and correlation dimensions respectively. }
$$
\sum_{i=1}^N p_i^q \asymp l^{\tau(q)}
\label{pq}
$$
As $l\to 0$, the integral in (\ref{falpha}) will be dominated by value of $\alpha$ which makes $q\alpha-f(\alpha)$ smallest and it follows that
$$
\tau(q)=\inf_\alpha\left[q\alpha-f(\alpha)\right] .
$$
Now we recall how to relate $f(\alpha)$-spectrum (\ref{falpha}) with the $\beta(q)$-spectrum (\ref{beta}): First, we divide the unit circle in the $w$-plane into segments of length $\epsilon$.  Let the number of segments that scale under mapping $z=F(w)$ as $\epsilon \to \epsilon^{\alpha}$ equals $\epsilon^{-\omega(\alpha)}$. Then
$$
\epsilon^{-\beta(q)} \asymp \int |F'\left(e^{\epsilon+\i\varphi}\right)|^qd\varphi \asymp \int \epsilon^{q(\alpha-1)}\epsilon^{-\omega(\alpha)}d\alpha
$$
and, therefore,
$$
\beta(q)=\sup_{\alpha}\left[\omega(\alpha)-(\alpha-1)q\right]
$$
The number of segments in the $w$ plane that scale under the mapping $z=F(w)$ as $\epsilon \to \epsilon^{\alpha}$ equals the number of covering boxes of size $l=\epsilon^\alpha $ in the $z$-plane that scale under the inverse mapping as $l \to l^{1/\alpha}$. Since harmonic measure is inversely proportional to $|F'(w)|$, we obtain that $l^{-f(1/\alpha)}=\epsilon^{-\omega(\alpha)}$, i.e. $\alpha f(1/\alpha)=\omega(\alpha)$. Thus
$$
\beta(q)=\sup_{\alpha}\left[q-1+(f(\alpha)-q)/\alpha\right]
$$
and, as a consequence
$$
f(\alpha)=\inf_q\left[q+\alpha(\beta(q)+1-q)\right] .
$$
Finally, note that the only scaling exponents $\alpha$ for which $f(\alpha)>0$ are present in the spectrum, because the negative $f$ correspond to zero probability events, i.e. for $f(\alpha)<0$, $l^{-f(\alpha)}\to 0$ as $l\to 0$.

\end{document}